# Complex Fluid-Fluid Interface may Non Trivially Dictate Droplet Deformation in an Incipient Flow

Sayan Das, Shubhadeep Mandal and Suman Chakraborty[*]

*Department of Mechanical Engineering, Indian Institute of Technology Kharagpur,
Kharagpur – 721302, India*

**Abstract**

The present study theoretically predicts the effect of interfacial viscosity on the deformation of a compound drop as well as on the bulk rheology. The system at hand comprises of a dilute emulsion of concentric compound drops, laden with surfactants and suspended in a linear flow. Two types of linear flows are considered in this study, namely, a uniaxial extensional flow and a simple shear flow. Presence of surfactants along the drop surface leads to the generation of an interfacial viscosity, which is different from the bulk. This interfacial viscosity generates a viscous drag that along with bulk flow-induced nonuniform surfactant distribution on the drop surface significantly alters drop dynamics. For the present study an asymptotic approach is used to solve the flow field under the limiting case of diffusion-dominated-surfactant transport. Assuming the surfactants to be bulk-insoluble and negligible inertia to be present in fluid flow, it is shown that presence of interfacial viscosity reduces the deformation of a compound drop and enhances the stability of a dilute double emulsion. At the same time the effective viscosity of the emulsion also increases with rise in interfacial viscosity. For large values of interfacial dilatational viscosity the drop deformation is seen to increase and hence the stability of the double emulsion is questionable.

[*]Corresponding author, email: suman@mech.iitkgp.ernet.in

## I. Introduction

Double emulsions or compound drops are multiphase systems comprising of a drop suspended in a second immiscible phase which as a whole is suspended in a third immiscible phase. In short, double emulsions can be stated as an 'emulsion of an emulsion'. Because of its special structure it can be used as a model to study vesicle or cellular dynamics in arteries or microchannels [1,2]. Double emulsions have a wide spectrum of applications, some of which are cosmetics, food and pharmaceutical industries [3,4], materials processing [4,5], hydrocarbon separation as well as waste water treatment [6–8]. This sort of system that consists of double emulsions physically exist during phase separation [9], drug delivery [10,11], lipid bilayer formation [12] and recovery of oil through porous structures [13]. Double emulsions have been an area of interest to different researchers since a long time. Available literatures in this field show that compound drops can be experimentally generated in microfluidic devices [14,15]. Li et al. [14] in their experimental study, generated monodisperse water/oil/water (W/O/W) compound drops with the help of three capillary tubes, while Pal [15] used re-emulsification of W/O or O/W drops in the presence of surfactants to generate compound drops.

Existing literatures show the effect of linear flows on the deformation characteristics of compound drops. Stone and Leal [13] showed, both theoretically as well as numerically, that the outer interface of a compound drop suspended in a uniaxial extensional flow deforms to a prolate shape whereas the inner drop deforms to an oblate spheroid due to biaxial flow in the annular region. The reverse is also found to be true when the bulk flow is a biaxial flow. For the case of a single drop, it was shown that the drop deformed to a prolate shape. Later Mandal et al. [16] showed the effect of surfactants on the deformation of a compound drop suspended in a linear flow. They theoretically predicted that the presence of surfactants leads to increase in deformation of the inner drop whereas the deformation of the outer drop solely depends on the radius ratio of the outer drop to that of the inner drop ($R$). For the case of a single drop $(R \to 0)$, the deformation was found to be the lowest. However, in the presence of surfactants, the deformation of both the inner and the outer drop surface was found to gradually increase with increase in $R$. They also showed that same characteristics in deformation is seen for both simple shear as well as uniaxial extensional flow. The effective viscosity of an emulsion of compound drops suspended in a linear flow was also found to increase due to presence of surfactants [16]. Various experimental studies performed till date have confirmed the presence of surfactants or surface active agents in different multiphase systems [15,17,18] in the form of impurities or as additives. Surfactants has been used as an additive in different studies to stabilize emulsions [17,18]. For example, in the experimental study performed by Pal [17], addition of surfactants have been solely for the purpose of achieving a stable double emulsion. A few studies available in the literature state the effect of surfactant distribution on the dynamics of compound drops [19–21,16,22–24]. Some of these studies include the effect of surfactants on the migration of a deformable compound drop in arbitrary imposed flow [16] and thermocapillary migration of surfactant-laden compound drops [20–23]. The rheology of an emulsion of drops is also affected due to presence of surfactants. Following the pioneering work done by Einstein [25] on the effective viscosity of an emulsion of particles, a number of studies were performed to investigate the bulk rheology of a suspension of particles or drops [26–32]. In the study done by Stone & Leal [13], it was shown that suspension comprising of compound drops effectively behaves as a suspension of rigid particles as the inner drop radius approached the outer one. It should also be noted that the empirical relation proposed by Masumoto and coworkers [33,34] can be used to predict the effective viscosity of double emulsions. On a different note, Pal [17] executed different experiments on the rheology of double emulsions with varying drop concentrations to show that the constitutive behavior of these emulsions depend strongly on the volume fraction as well as the viscosity of the drops.

It has been previously shown that the presence of surfactants along interfaces result in an interfacial viscosity which is different from the bulk [35–37]. The surfactants thus play a dual role; on one hand it induces Marangoni stress by varying the surface tension along the drop surface [38,39], while on the other hand it generates surface-excess viscous stresses [40–42]. Different researchers have explored the effect of interfacial shear as well as dilatational viscosities on the dynamics of drops in the presence of imposed linear and Poiseuille flows, temperature gradient and gravity [36,42,43]. In the experimental study performed by Pal [17], surfactants were used to stabilize an emulsion of compound drops. However, they did not take into account the interfacial viscosity which is generated due to presence of surfactants along the drop surface. Hence, in the present study, we develop a more realistic model of a double emulsion in the presence of surfactant-induced interfacial viscosity.

The present study is performed keeping the following objectives in mind: to study the effect of interfacial viscosity on (i) the deformation behavior of an isolated compound drop, and (ii) on the bulk rheology of a dilute emulsion comprising of compound drops. According to previous studies [36,44], the fluid-fluid interface is modeled as a two-dimensional Newtonian fluid possessing constant shear and dilatational surface viscosities. The surface-excess viscous stress generated is assumed to obey the Boussinesq-Scriven constitutive law [36]. One of the important findings in this study is that interfacial viscosity-induced surface viscous stresses result in a reduction in the deformation of a compound drop and hence play a significant role in stabilizing their dilute emulsions. It is also predicted that interfacial viscosity significantly alters the suspension rheology by increasing the effective viscosity of such dilute emulsions.

## II. Problem formulation

### A. Physical system

The present system consists of a dilute suspension of concentric drops as shown in the schematic in Figure 1. In this 'compound drop' system considered, the outer drop has a radius of $R_o$ while the inside drop has a radius of $R_i$. Both the drop phases as well the carrier phase are Newtonian. We denote the carrier phase by 1, the outer drop phase by 2 and the inner drop phase by 3. Different fluidic properties are denoted by using either of these numbers as subscripts. For instance, the density of the *i*th fluid is denoted by $\rho_i$, and in a similar manner the bulk viscosity of the *j*th fluid is denoted by $\mu_j$. All the interfacial properties are denoted by subscript '*ij*'. The interfacial tension between the *i*th and the *j*th fluid is denoted by $\bar{\sigma}_{ij}$. Bulk-insoluble surfactants are considered to be present at the interface, which only get transported along the drop surface. The surfactant concentration at *ij*th interface is denoted by $\bar{\Gamma}_{ij}$. In the absence of fluid flow, the equilibrium surfactant concentration is $\bar{\Gamma}_{ij}^{eq}$, the corresponding equilibrium surface tension is $\bar{\sigma}_{ij}^{eq}$. The equilibrium surface tension for a clean surface is indicated by $\bar{\sigma}_{ij}^{c}$. Presence of surfactants in general leads to the generation of an interfacial viscosity, different from the bulk [36]. This interfacial viscosity may be shear or dilatational, which generates surface-excess viscous stresses that affect the drop dynamics. Our prime focus in this study is to investigate the effect of these surface-excess viscous stresses on the deformation as well as suspension rheology of a double emulsion. The shear and dilatational viscosities along the $ij$th interface are denoted by $\mu_{ij}^s$ and $\mu_{ij}^d$ respectively. The $ij$th interface is itself denoted by $S_{ij}$. The suspending phase is considered to undergo a linear flow, that may be either a uniaxial extensional flow or a simple shear flow. The bulk linear flow, $\bar{\mathbf{u}}_\infty$, in its general form can be represented as

$$\bar{\mathbf{u}}_\infty = \left(\bar{\mathbf{D}}_\infty + \bar{\mathbf{\Omega}}_\infty\right) \cdot \bar{\mathbf{x}}, \tag{1}$$

where $\bar{\mathbf{x}}$ is the position vector, $\bar{\mathbf{D}}_\infty$ is the rate of strain tensor and $\bar{\mathbf{\Omega}}_\infty$ is the vorticity tensor. For the case of a uniaxial extensional flow we have

$$\bar{\mathbf{D}}_\infty = \frac{\dot{\gamma}}{2}\begin{bmatrix} -1 & 0 & 0 \\ 0 & -1 & 0 \\ 0 & 0 & 2 \end{bmatrix}, \quad \bar{\mathbf{\Omega}}_\infty = \mathbf{0}, \tag{2}$$

where $\dot{\gamma}$ is the extension rate. On the other hand for the case of a simple shear flow we have

$$\bar{\mathbf{D}}_\infty = \frac{\dot{\gamma}}{2}\begin{bmatrix} 0 & 1 & 0 \\ 1 & 0 & 0 \\ 0 & 0 & 0 \end{bmatrix}, \quad \bar{\mathbf{\Omega}}_\infty = \frac{\dot{\gamma}}{2}\begin{bmatrix} 0 & 1 & 0 \\ -1 & 0 & 0 \\ 0 & 0 & 0 \end{bmatrix}, \tag{3}$$

where $\dot{\gamma}$ is the rate of shear. The initial uniform distribution of surfactants along the interface of either of the drops $\left(\bar{\Gamma}_{ij}^{eq}\right)$ is disturbed in the presence of an imposed linear flow. The resulting nonuniform distribution of surfactants generates a gradient in surface tension along both the inner and outer interfaces ($S_{12}$ and $S_{23}$) which gives birth to a Maranogoni stress. This Marangoni stress accompanied by the bulk viscous stress and surface-excess viscous stress (shear and dilatational) lead to the deformation of the drop. For the present problem, we use a spherical coordinate system $(\bar{r}, \theta, \varphi)$ which is attached to the origin of outer drop, that is, at the centroid of the compound drop system (see Figure 1).

## B. Important assumptions

Towards modeling a physically realistic system, we invoke the following assumptions
(i) The surface tension force as well as the viscous and the pressure forces are assumed to be much larger as compared to the inertia forces in fluid flow. Thus the present study falls under the creeping flow regime and hence Reynolds number in each of the phases is assumed to be much less than unity.
(ii) The surface tension force acting along the interfaces is assumed to dominate the viscous forces, such that only small deformation of either of the drops are taken into consideration. Thus the capillary number, $Ca$, which is the governing parameter for drop deformation is taken to be small $\left(Ca^* = \mu_1 \dot{\gamma} R_o / \bar{\sigma}_{12}^c \ll 1\right)$.
(iii) The transport of surfactants is assumed to take place along the drop surface only.
(iv) The double emulsion drop system is assumed to be stable. Some recent experimental studies by Utada et al. [45], Fichuex et al. [46], Pal et al. [17] and Kim et al. [47] show that stable configurations can be generated even though the thickness of the annular region in between the outer and the inner drop phase is small. In addition to this, presence of surfactants enhances the stability of emulsions [25].

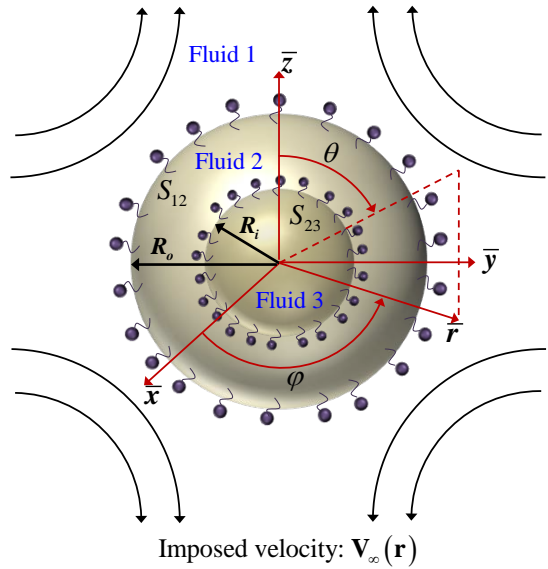

Figure 1. Schematic of a surfactant-laden compound drop suspended in a linear flow field. The outer as well as inner drops are concentric with radius $R_o$ and $R_i$ respectively. As an example we have shown the imposed flow to be a uniaxial extensional flow. Both the cartesian $(\bar{x}, \bar{y}, \bar{z})$ as well as the spherical coordinates $(\bar{r}, \theta, \varphi)$ are shown to be attached to the centroid of the outer drop. Surfactants are distributed at the interface of either of the drops.

(v) We assume a linear relationship to exist between interfacial tension and the surfactant distribution along the drop surface.
(vi) The above system is taken to be unbounded. In other words, the any effect of bounding wall on the drop is neglected by considering the outer drop radius to be significantly small with respect to the channel height.
(vii) The fluid-fluid interfaces are assumed to behave as a two dimensional Newtonian fluid, with constant shear and dilatational viscosities, that obeys the Boussinesq-Scriven constitutive law.
Experimental relevance of the above assumptions is provided at a later section.

## C. Governing equations and boundary conditions
We now state the dimensional form of the governing equations and boundary conditions for the present problem. The fluid flow of the $i$th fluid $(i = 1, 2, 3)$ under the assumption of negligible inertial effects is governed by the Stokes equation which can be stated as
$$\bar{\nabla}\bar{p}_i = \mu_i \bar{\nabla}^2 \bar{\mathbf{u}}_i, \quad \bar{\nabla} \cdot \bar{\mathbf{u}}_i = 0, \tag{4}$$
where $(\bar{\mathbf{u}}_i, \bar{p}_i)$ is the velocity and the pressure field of the $i$th phase. The above flow field equation is subjected to the following far-field boundary condition

$$\text{at } \bar{r} \to \infty, \quad (\bar{\mathbf{u}}_1, p_1) = (\bar{\mathbf{u}}_\infty, \bar{p}_\infty), \tag{5}$$

Also the velocity field at the centroid of compound drop is bounded. The other boundary conditions at the interface of the drop are mentioned below

(i) The kinematic boundary condition indicates that the normal component of velocity field at either of the interfaces is zero, that is

$$\text{at } S_{ij}, \quad \bar{\mathbf{u}}_i \cdot \mathbf{n}_{ij} = \bar{\mathbf{u}}_j \cdot \mathbf{n}_{ij} = 0, \tag{6}$$

where $\mathbf{n}_{ij}$ is the normal drawn to '$ij$'th interface $S_{ij}$ and can be expressed as $\mathbf{n}_{ij} = \nabla F_{ij} / |\nabla F_{ij}|$, $F_{ij}$ being the equation of the $ij$th surface.

(ii) The continuity in tangential component of velocity field at interface $S_{ij}$ along with the kinematic condition results in another boundary condition given below

$$\bar{\mathbf{u}}_i = \bar{\mathbf{u}}_j. \tag{7}$$

(iii) Finally the stress balance condition at the $ij$th interface is given by

$$\left(\bar{\boldsymbol{\tau}}_i \cdot \mathbf{n}_{ij} - \bar{\boldsymbol{\tau}}_j \cdot \mathbf{n}_{ij}\right) = \bar{\sigma}_{ij}\left(\bar{\nabla} \cdot \mathbf{n}_{ij}\right)\mathbf{n}_{ij} - \bar{\nabla}_s \bar{\sigma}_{ij} - \bar{\nabla}_s \cdot \bar{\boldsymbol{\tau}}_{s,ij}, \tag{8}$$

where $\boldsymbol{\tau}_i = -\bar{p}_i \mathbf{I} + \mu_i \left[\bar{\nabla}\bar{\mathbf{u}}_i + \left(\bar{\nabla}\bar{\mathbf{u}}_i\right)^T\right]$ is the stress tensor in the $i$th fluid consisting of both hydrostatic and deviatoric parts and $\nabla_s = \left(\mathbf{I} - \mathbf{n}_{ij}\mathbf{n}_{ij}^T\right) \cdot \nabla$ is the surface gradient operator. The last term on the right hand side of equation (8) denotes the traction developed due to surface-excess viscous stress along the interface $S_{ij}$. In accordance to the Boussinesq-Scriven constitutive law for Newtonian fluids, we can express the surface-excess viscous stress $\left(\bar{\boldsymbol{\tau}}_{s,ij}\right)$ along the $ij$th interface in the following form [36]

$$\bar{\boldsymbol{\tau}}_{s,ij} = 2\mu_{ij}^s \bar{\mathbf{D}}_s + \left(\mu_{ij}^d - \mu_{ij}^s\right)\{\mathbf{I}_t : \bar{\mathbf{D}}_s\}\left(\mathbf{I} - \mathbf{n}_{ij}\mathbf{n}_{ij}^T\right), \tag{9}$$

where $\bar{\mathbf{D}}_s$ denotes the surface rate of deformation tensor and can be represented as

$$\bar{\mathbf{D}}_s = \frac{1}{2}\left\{\bar{\nabla}_s \bar{\mathbf{u}} \cdot \left(\mathbf{I} - \mathbf{n}_{ij}\mathbf{n}_{ij}^T\right) + \left(\mathbf{I} - \mathbf{n}_{ij}\mathbf{n}_{ij}\right) \cdot \left(\bar{\nabla}_s \bar{\mathbf{u}}\right)^T\right\}. \tag{10}$$

In the above expressions, subscript '$s$' indicates a surface quantity. The first and second term on the RHS of the equation (8) are the traction developed due to drop surface curvature and Marangoni stress generated due to nonuniform surfactant distribution on the surface of the drop.

The surfactant distribution is governed by a convection-diffusion equation which, at the $ij$th interface can be written in the following form [25]

$$\bar{\nabla}_s \cdot \left(\bar{\mathbf{u}}_{s,ij}\bar{\Gamma}_{ij}\right) = D_{ij}\bar{\nabla}_s^2 \bar{\Gamma}_{ij}, \tag{11}$$

where $\bar{\mathbf{u}}_{s,ij}$ and $D_{ij}$ represents the fluid velocity and surface diffusivity of the surfactants at the $ij$th interface. The above equation is also known as the surfactant transport equation. The left hand side of the above equation signifies surfactant transport along the interface $S_{ij}$ due to surface convection and that on the right denotes surfactant transport due to surface diffusion. Finally the equation of state that linearly relates the interfacial tension with the surfactant distribution reads as [25]

$$\bar{\sigma}_{ij} = \bar{\sigma}_{ij}^c - \bar{\Gamma}_{ij} R_g T \tag{12}$$

We next proceed towards obtaining the dimensionless set of governing equations and boundary conditions. Towards this, the following characteristic scales are used

$$u \sim \dot{\gamma}R_o, r \sim R_o, p_c \sim \tau_c \sim \mu_1\dot{\gamma}, \bar{\sigma}_{ij} \sim \bar{\sigma}_{12}^c, \Gamma_{12} \sim \bar{\Gamma}_{12}^{eq}, \Gamma_{23} \sim \bar{\Gamma}_{23}^{eq}, D_{ij} \sim D_{12}, \qquad (13)$$

It has to be noted that the quantities with an 'overbar' denote dimensional quantities and that without any 'overbar' represent dimensionless ones. Upon utilization of the above non-dimensional scheme, some important property ratios as well as non-dimensional numbers crop up while deriving the dimensionless form of the governing equations and boundary conditions. The important ratios that are involved in the present analysis are $\lambda_i = \mu_i/\mu_1$, $d = D_{23}/D_{12}$, $\eta = \bar{\Gamma}_{23}^{eq}/\bar{\Gamma}_{12}^{eq}$, $R = R_i/R_o$ and $\xi = \bar{\sigma}_{23}^c/\bar{\sigma}_{12}^c$. Thus the governing equation for flow-field in its non-dimensional form is given by

$$-\nabla p_i + \lambda_i \nabla^2 \mathbf{u}_i = \mathbf{0}, \quad \nabla \cdot \mathbf{u}_i = 0, \quad \text{where } \lambda_i = \delta_{i1} + \lambda_i \delta_{i2} + \lambda_i \delta_{i3}, \qquad (14)$$

where $\delta_{ij}$ is the Kronecker delta. The above governing equations are subjected to the following boundary conditions at interface $S_{ij}$

$$\left.\begin{array}{l}\mathbf{u}_i \cdot \mathbf{n}_{ij} = \mathbf{u}_j \cdot \mathbf{n}_{ij} = \mathbf{0}, \\ \mathbf{u}_i = \mathbf{u}_j, \\ \boldsymbol{\tau}_i \cdot \mathbf{n}_{ij} - \boldsymbol{\tau}_j \cdot \mathbf{n}_{ij} = \dfrac{1}{Ca^*}\left\{\bar{\sigma}_{ij}\left(\nabla_s \cdot \mathbf{n}_{ij}\right)\mathbf{n}_{ij} - \nabla_s \bar{\sigma}_{ij}\right\} - \nabla_s \cdot \boldsymbol{\tau}_{s,ij},\end{array}\right\} \qquad (15)$$

where the traction due to surface-excess viscous stress along the interface $S_{ij}$ can be expressed as follows [36]

$$\nabla_s \cdot \boldsymbol{\tau}_{s,ij} = 2Bo_{ij}^{(s)}\left\{(\mathbf{I}_t \cdot \nabla) \cdot (\mathbf{I}_t \cdot \mathbf{D}_s \cdot \mathbf{I}_t)\right\} + \left(Bo_{ij}^{(d)} - Bo_{ij}^{(s)}\right)\left\{\mathbf{I}_t \cdot \nabla(\mathbf{I}_t : \nabla \mathbf{u}_i) + 2H(\mathbf{I}_t : \nabla \mathbf{u}_i)\right\}, \qquad (16)$$

where $\mathbf{I}_t = \mathbf{I} - \mathbf{n}_{ij}\mathbf{n}_{ij}^T$ is the surface projection tensor and $H$ is the mean curvature which is equal to -1 for a spherical drop. $Bo_{ij}^{(s)}$ and $Bo_{ij}^{(d)}$ are dimensionless shear and dilatational Boussinesq numbers for the interface $S_{ij}$ respectively. They physically signify the relative strength of the interfacial shear or dilatational surface viscous stresses with respect to the bulk viscous stresses acting on the drop. They can be expressed as follows

$$Bo_{ij}^{(s)} = \frac{\mu_{ij}^s}{\mu_e R_o}, \quad Bo_{ij}^{(d)} = \frac{\mu_{ij}^d}{\mu_e R_o}. \qquad (17)$$

The dimensionless form of the surface-excess viscous stress is given by [44]

$$\boldsymbol{\tau}_{s,ij} = \left[Bo_{ij}^{(d)} - Bo_{ij}^{(s)}\right](\nabla_s \cdot \mathbf{u}_s)\mathbf{I}_t + 2Bo_{ij}^{(s)}\mathbf{D}_s, \qquad (18)$$

The non-dimensional surfactant transport equation, governing the distribution of surfactants along either of the interfaces, $S_{12}$ and $S_{23}$, is given below [16]

$$\frac{Pe_s}{d_{ij}}\nabla_s \cdot (\mathbf{u}_{s,ij}\Gamma_{ij}) = \nabla_s^2 \Gamma_{ij}, \quad \text{where } d_{ij} = \delta_{i1}\delta_{j2} + d\delta_{i2}\delta_{j3}. \qquad (19)$$

In the above set of governing equations and boundary conditions, we encounter the non-dimensional number $Pe_s = \dot{\gamma}R_o^2/D_{12}$, which is the surface Péclet number and signifies the relative importance of surfactant transport along the drop interface $S_{12}$ due to convection as compared to diffusion. The non-dimensional equation of state at the two interfaces, $S_{12}$ and $S_{23}$, are defined in the following manner

$$\sigma_{12} = 1 - \beta\Gamma_{12}, \quad \sigma_{23} = \xi - \eta\beta\Gamma_{23}, \qquad (20)$$

where $\beta = \bar{\Gamma}_{12}^{eq} RT / \bar{\sigma}_{12}^{c}$ is the elasticity parameter [48], which signifies the sensitivity of surfactant concentration towards the interfacial tension of the outer drop. The value of $\beta$ varies between 0 and 1. Following Stone & Leal [18], we define a modified capillary number as $Ca = Ca^{*}/(1-\beta)$ which can be expressed as $Ca = \mu_1 \dot{\gamma} R_o / \bar{\sigma}_{12}^{eq}$, where $\bar{\sigma}_{12}^{eq} = (1-\beta)\bar{\sigma}_{12}^{c}$ is the equilibrium surface tension of a surfactant-laden drop. Thus the modified capillary number is defined based on $\bar{\sigma}_{12}^{eq}$ rather than $\bar{\sigma}_{12}^{c}$ (surface tension for a clean drop). In addition to the surfactant transport equation mentioned in equation (19), the surfactant concentration should also satisfy the constraint on mass conservation along the interface which can be expressed as [16,49]

$$\int_{\varphi=0}^{2\pi} \int_{\theta=0}^{\pi} \Gamma_{ij}(\theta,\varphi) r_{s,ij}^{2} \sin\theta \, d\theta \, d\varphi = 4\pi. \tag{21}$$

where $r_{s,ij}$ is the radial position of the $ij$th interface. A close look into the equations (14)-(19) clearly suggests that the governing equations for flow field, although simplified, are still coupled with the surfactant transport equation through the convection term. Consideration of the deformability of the drop further brings in nonlinearity in the system of equations. Thus the present system of governing equations cannot be solved for arbitrary values $Ca$ or $Pe_s$. However, solution of such a system of equations can be obtained with the help of asymptotic analysis. In the present study we make use of the domain perturbation method to explore two different limiting cases of surfactant transport, namely, a low surface Péclet number limit ($Pe_s \ll 1$) that signifies that surfactant transport occurs mainly due to surface diffusion rather surface convection and a high surface Péclet number limit ($Pe_s \to \infty$) which indicates that the primary mode of surfactant transport along the drop surface is due to surface convection. In the present scenario as only small deviations from the spherical shape of the drop is assumed, so low values of $Ca$ has to be used.

## III. Asymptotic analysis

### A. Low surface Péclet number limit ($Pe_s \ll 1$)

In the low surface Péclet number limit, the surfactant transport along the drop surface is dominated by surface diffusion in comparison to surface advection. The magnitude of both $Pe_s$ as well as $Ca$ are low in this limit and hence we make a simple assumption that $Pe_s \sim Ca$ [48] or it can be written in the following form

$$Pe_s = kCa, \tag{22}$$

where $k = R_o \bar{\sigma}_{12}^{c}(1-\beta)/\mu_1 D_{12}$ is known as the property parameter and has a magnitude of the order of 1. Thus shape deformation of the drop is solely a function of $Ca$ provided the values of $k$ and $\beta$ are given. Hence we choose $Ca$ as the perturbation parameter in our asymptotic analysis. Thus any flow variable $\psi$ can be expanded in an increasing power series of $Ca$ in the following manner

$$\psi = \psi^{(0)} + \psi^{(Ca)} Ca + O(Ca^2), \tag{23}$$

where the first term in the above expression is the leading-order term that corresponds to the spherical shaped drop. The second term on the right side of the above equation is a correction term due to the shape deformation of the drop. The surfactant concentration depend on the mode of its transport along the drop surface and hence it can be expanded as

$$\Gamma_{ij} = 1 + kCa\Gamma_{ij}^{(1)} + O(Ca^2), \tag{24}$$

where $\Gamma_{ij}^{(1)}$ denotes the nonuniform surfactant concentration along the interface $S_{ij}$ due to O($Ca$) shape deformation of the drop. At each order of perturbation the surfactant concentration must satisfy the mass conservation constraint [see equation (21)] along the interface $S_{ij}$. The local surfactant concentration being independent of $r$, it can be expressed in terms of spherical surface harmonics as follows

$$\Gamma_{ij} = \sum_{n=0}^{\infty}\sum_{m=0}^{n}\left[\left(\Gamma_{ij}\right)_{n,m}\cos(m\varphi) + \left(\hat{\Gamma}_{ij}\right)_{n,m}\sin(m\varphi)\right]P_{n,m}(\cos\theta), \tag{25}$$

where $P_{n,m}(\cos\theta)$ is the associate Legendre polynomial of degree $n$ and order $m$. To obtain the surfactant concentration along the drop surface, we have to calculate the constant coefficients $\left(\Gamma_{ij}\right)_{n,m}$ and $\left(\hat{\Gamma}_{ij}\right)_{n,m}$, which can be obtained from the surfactant transport equation. The deformed drop surface, $S_{ij}$, in spherical coordinates, can be expressed in the following manner

$$F_{ij} = r - R_{ij}\left\{1 + Ca g_{ij}^{(Ca)} + O(Ca^2)\right\}, \text{ where } R_{ij} = \delta_{i1}\delta_{j2} + R\delta_{i2}\delta_{j3}, \tag{26}$$

where $g_{ij}^{(Ca)}(\theta,\varphi)$ is the $O(Ca)$ correction to the drop shape due to the combined effect of imposed linear flow and its resulting surfactant redistribution. Similar to equation (25), $g_{ij}^{(Ca)}$ can be expressed as

$$g_{ij}^{(Ca)} = \sum_{n=0}^{\infty}\sum_{m=0}^{n}\left[\left(L_{ij}^{(Ca)}\right)_{n,m}\cos(m\varphi) + \left(\hat{L}_{ij}^{(Ca)}\right)_{n,m}\sin(m\varphi)\right]P_{n,m}(\cos\theta), \tag{27}$$

where $\left(L_{ij}^{(Ca)}\right)_{n,m}$ and $\left(\hat{L}_{ij}^{(Ca)}\right)_{n,m}$ are unknown constant coefficients. Since $Ca \ll 1$, we solve only for the leading-order flow field and surfactant concentration, that is $\mathbf{u}_i^{(0)}, p_i^{(0)}, \Gamma_{ij}^{(1)}$ and also the corresponding $O(Ca)$ shape deformation of the drop, $g_{ij}^{(Ca)}$. Substituting equations (22)-(24) in the non-dimensional governing equations and its respective boundary conditions and comparing equal coefficients of $Ca$ on either sides of the equation, we obtain the leading-order governing equations and boundary conditions. As we have no intension of obtaining any higher order solutions of the flow field or surfactant concentration, we ignore the superscript '(0)' ['(1)' for surfactant concentration] in the subsequent leading-order expressions. The leading-order flow field governing equations in its dimensionless form is given by equation (14). The leading-order surfactant transport equation on the interfaces is given below

$$\frac{1}{d_{ij}}\nabla_s \cdot (\mathbf{u}_{ij}) = \nabla_s^2 \Gamma_{ij}. \tag{28}$$

The relevant leading-order boundary conditions at the interface $S_{ij}$ are as follows

$$\left.\begin{array}{l}\mathbf{u}_i\big|_{S_{ij}}\cdot\mathbf{n}_{ij}=\mathbf{u}_j\big|_{S_{ij}}\cdot\mathbf{n}_{ij}=\mathbf{0},\\[4pt]
\mathbf{u}_i\big|_{S_{ij}}=\mathbf{u}_j\big|_{S_{ij}},\\[4pt]
\left[\boldsymbol{\tau}_i\big|_{S_{ij}}\cdot\mathbf{n}_{ij}-\boldsymbol{\tau}_j\big|_{S_{ij}}\cdot\mathbf{n}_{ij}\right]\cdot\left(\mathbf{I}-\mathbf{n}_{ij}\mathbf{n}_{ij}^T\right)=\eta_{ij}\zeta_2\nabla_s\Gamma_{ij}-\left(\nabla_s\cdot\boldsymbol{\tau}_{s,ij}\right)\cdot\left(\mathbf{I}-\mathbf{n}_{ij}\mathbf{n}_{ij}^T\right),\\[4pt]
\left[\boldsymbol{\tau}_i\big|_{S_{ij}}\cdot\mathbf{n}_{ij}-\boldsymbol{\tau}_j\big|_{S_{ij}}\cdot\mathbf{n}_{ij}\right]\cdot\mathbf{n}_{ij}=\dfrac{1}{R_{ij}}\left(\dfrac{2\zeta_{1,ij}}{Ca}-2\zeta_{1,ij}g_{ij}-\zeta_{1,ij}\nabla_s^2 g_{ij}-2\eta_{ij}\zeta_2\Gamma_{ij}\right)-\left(\nabla_s\cdot\boldsymbol{\tau}_{s,ij}\right)\cdot\mathbf{n}_{ij},\end{array}\right\}\quad(29)$$

where $\eta_{ij}=\delta_{i1}\delta_{j2}+\eta\delta_{i2}\delta_{j3}$, $\zeta_{1,ij}=\delta_{i1}\delta_{j2}+\zeta_1\delta_{i2}\delta_{j3}$. In equation (29), some dimensionless constants appear whose expressions are given below

$$\zeta_1=\dfrac{\xi-\eta\beta}{1-\beta},\quad \zeta_2=\dfrac{\beta k}{1-\beta}. \qquad (30)$$

The boundary conditions in equation (29) consist of the leading-order no-slip boundary condition, the kinematic boundary condition, the shear stress boundary condition and lastly the normal stress balance. Other than the last one, all other boundary conditions are used to obtain the leading-order flow field. The last boundary condition, which is the normal stress balance, helps us find out the $O(Ca)$ deformation of both the interfaces, $S_{12}$ and $S_{23}$. As stated before the range of $\beta$ can be expressed as $0\leq \beta \leq 1$. Irrespective of the surfactant concentration, the surface tension along the inner interface, $S_{23}$ is always positive. Since at equilibrium, $\sigma_{23}^{eq}=\xi-\eta\beta$, we can thus write that $\xi\gg\eta\beta$. We next proceed towards solving the above set of governing equations that are subjected to relevant boundary conditions. We make use of the Lamb's general solution as was previously used by Haber & Hetsroni [50] for a single drop as well as that by Mandal et al. [16], where the effect of surfactant Marangoni stress on the dynamics of a compound drop was analyzed. The details of Lamb's general solution for the flow field of all the phases are provided in section A of the supplementary material. The far field for the present problem may be either a simple shear flow or a uniaxial extensional flow. We first utilize the flow field boundary conditions as well as the surfactant transport equation at both the interfaces to solve for the velocity and pressure field and the surfactant concentration. Then using the normal stress balance on either of the interfaces, we obtain the drop shape.

### 1. Experimental Relevance

We now discuss the experimental relevance of the assumptions made in the present analysis. Experiments dealing with the low Reynolds number hydrodynamics of both a single drop [51,52] as well as a compound drop [17,53] has been performed previously. For instance, Tsukada et al. [53] have performed experiments on compound drops with vegetable oil as the outer drop phase and silicone oil as the inner drop as well as the carrier phase. The viscosity of vegetable oil is 0.254 Pa-s and the interfacial tension at room temperature between the carrier phase and the outer drop phase is $\sigma_{12}=0.003$ N/m. Thus for drop size of the order of $R_o\sim 10^{-3}$ m with its characteristic shear (extensional rate) $\dot{\gamma}=1\text{s}^{-1}$, we see that the capillary number is $Ca=\mu_1\dot{\gamma}R_o/\sigma_{12}\sim 0.08$ which is consistent with our assumption of a nearly spherical drop, but with the presence of small deformation. According to the experiments performed by Stebe et al. [54], the typical range of values for the surface diffusivity of surfactants is $D_{12}\sim 10^{-6}$-$10^{-9}$ m$^2$/s.

The order of magnitude of k = $Pe_s/Ca \sim O(1)$ for the above property values, which is again consistent with our assumption of low surface Péclet number. As the surface diffusivity of surfactants varies over a wide range depending on the surfactant-fluid interaction, the order of magnitude of $k$ may rise up to $k \sim O(100)$ representing high $Pe_s$ limit. These range of higher values of $k$ was previously used by Stone & Leal [18]. Similar asymptotic limits were as well used by Mandal et al. [55] and a good match between the asymptotic and numerical simulations were found.

## 2. Leading-order solution

The leading-order solution that is presented below is for the special case of a concentric compound drop. In this section we highlight the important results, namely the O(*Ca*) shape deformation of the drop and the surfactant concentration along both the interfaces which are obtained as a solution of the coupled flow-field and convection-diffusion equation for surfactant transport. The governing equations and boundary conditions are solved asymptotically using the same strategy as was followed by Mandal et al. [16]. A brief discussion of the present asymptotic approach as well as the general expression of the velocity in all the three phases is provided in section A of the supplementary material.

We now present the important results for two instances of linear flows, the uniaxial extensional flow and the simple shear flow at both the interfaces, $S_{12}$ and $S_{23}$.

### Uniaxial extensional flow

A schematic of a compound drop suspended in a uniaxial extensional flow has been already shown in the Figure 1. The surfactant concentration at both the interfaces ($S_{12}$, $S_{23}$) in the presence of an imposed uniaxial extensional flow is given by equation (25), where the only non-zero coefficients are provided below

$$\left(\Gamma_{ij}\right)_{2,0} = \Pi_{ext}^{(ij)} + \underbrace{\Xi_{ext}^{(ij)}}_{\text{Contribution of interface viscosity}}, \tag{31}$$

where $\Pi_{ext}^{(ij)} \equiv \Pi_{ext}^{(ij)}\left(\lambda_2, \lambda_3, R, d, \eta, \zeta_2\right)$ and $\Xi_{ext}^{(ij)} \equiv \Xi_{ext}^{(ij)}\left(Bo_{12}^{(s)}, Bo_{12}^{(d)}, Bo_{23}^{(s)}, Bo_{23}^{(d)}, \lambda_2, \lambda_3, R, d, \eta, \zeta_2\right)$.

The first term on the right hand side of the above equation $\Pi_{ext}^{(ij)}$ is the same as was obtained by Mandal et al. [16]. This term don't take into consideration any effect of interface viscosity, whereas the other term, $\Xi_{ext}^{(ij)}$ denote the change in surfactant concentration at the interfaces, $S_{12}$, $S_{23}$ due to the presence of interfacial shear and dilatational viscosity. The subscript '*ext*' denotes imposed uniaxial extensional flow. The leading-order solution of the flow field is not provided here to preserve the conciseness. Due to the enormous size of the expression of $\Xi_{ext}^{(ij)}$ both of them are kept in their functional form in equation (31). The detailed expressions of the quantities in equation (31) are provided in a Matlab script file (.m file) in the supplementary material. It can be seen from the expression in equation (31), that the dilatational as well as the shear viscosity at the interface of the inner drop affects the surfactant concentration at the outer drop surface and vice versa. Thus it can be inferred that the above results are not obtained by a mere superposition.

The resulting deformed shape of both the inner as well as the outer drop due to the combined effect of the imposed flow, surfactant redistribution and interfacial viscosity is given by

$$r_{s,ij} = R_{ij}\left\{1 + Ca\, g_{ij}^{(Ca)} + O(Ca^2)\right\}, \tag{32}$$

where $g_{ij}^{(Ca)}$ is given by equation (27). The only non-zero constant coefficients in the expression for $g_{ij}^{(Ca)}$ can be expressed as

$$\left(L_{ij}^{(Ca)}\right)_{2,0} = \Omega_{ext}^{(ij)} + \underbrace{\Theta_{ext}^{(ij)}}_{\substack{\text{Contribution of}\\\text{interface viscosity}}}, \tag{33}$$

where $\Omega_{ext}^{(ij)} \equiv \Omega_{ext}^{(ij)}(\lambda_2, \lambda_3, R, d, \eta, \zeta_1, \zeta_2)$, $\Theta_{ext}^{(ij)} \equiv \Theta_{ext}^{(ij)}\left(Bo_{12}^{(s)}, Bo_{12}^{(d)}, Bo_{23}^{(s)}, Bo_{23}^{(d)}, \lambda_2, \lambda_3, R, d, \eta, \zeta_1, \zeta_2\right)$ and superscript '$ij$' denotes interface $ij$. In the above expression, the expression for $\Omega_{ext}^{(ij)}$ is the same as was obtained by Mandal et al. [16]. The contribution of interfacial viscosity, shear and dilatational, is due to $\Theta_{ext}^{(ij)}$. Again the detailed expressions are not given in the present text to preserve its conciseness. However, the same has been provided in the supplementary material in a Matlab script file (.m file). For the particular case of a single drop ($R \to 0$), we have the following

$$\lim_{R \to 0}\left(L_{12}^{(Ca)}\right)_{2,0} = \frac{5}{8}\frac{16 + 4\zeta_2 + 20Bo_{12}^{(s)} + 19\lambda_2}{5 + 9Bo_{12}^{(d)} + 5\lambda_2 + \zeta_2 + 10Bo_{12}^{(s)}}, \tag{34}$$

The deformation of the compound drop is measured in terms of a deformation parameter which can be expressed as follows

$$D_{ext} = \frac{r_{ext}(\theta = 0) - r_{ext}(\theta = \pi/2)}{r_{ext}(\theta = 0) + r_{ext}(\theta = \pi/2)} \tag{35}$$

where $r_{ext} = R_{ij}\left\{1 + Ca\, g_{ij}^{(Ca)}\right\}$ is the radial positions of the interface (outer or inner) of the deformed compound drop.

**Simple shear flow**

For the case of a simple shear flow similar results are obtained. The surfactant concentration along both the interfaces, $S_{12}$, $S_{23}$, can be expressed in a similar manner as was done for the case of a uniaxial extensional flow. The only non-zero constant coefficients present in the general expression for surfactant concentration (equation (25)) is provided below

$$\left(\Gamma_{ij}\right)_{2,1} = \Pi_{shear}^{(ij)} + \underbrace{\Xi_{shear}^{(ij)}}_{\substack{\text{Contribution of}\\\text{interface viscosity}}}, \tag{36}$$

where $\Pi_{shear}^{(ij)} \equiv \Pi_{shear}^{(ij)}(\lambda_2, \lambda_3, R, d, \eta, \zeta_2)$, $\Xi_{shear}^{(ij)} \equiv \Xi_{shear}^{(ij)}\left(Bo_{12}^{(s)}, Bo_{12}^{(d)}, Bo_{23}^{(s)}, Bo_{23}^{(d)}, \lambda_2, \lambda_3, R, d, \eta, \zeta_2\right)$ and subscript '*shear*' denotes simple shear flow. The details of the above result has been provided in the supplementary material (as Matlab script files). The leading-order flow field for the case of a simple shear flow is not reproduced here for the sake of brevity. The corresponding shape deformation of the drop is next obtained. Fortunately it turns out that the constant non-zero coefficients, in the expression for shape deformation [equation (27)] for the case of a simple shear flow are related to that for the case of a uniaxial extensional flow in the following manner

$$\left(L_{ij}^{(Ca)}\right)_{2,1}\bigg|_{shear\ flow} = \frac{1}{3}\left(L_{ij}^{(Ca)}\right)_{2,0}\bigg|_{extensional\ flow}, \tag{37}$$

The deformation parameter for the case of an imposed shear flow is given by

$$D_{shear} = \frac{\max\{r_{shear}(\theta=\pi/2,\varphi)\} - \min\{r_{shear}(\theta=\pi/2,\varphi)\}}{\max\{r_{shear}(\theta=\pi/2,\varphi)\} + \min\{r_{shear}(\theta=\pi/2,\varphi)\}}, \quad (38)$$

where $r_{shear} = R_{ij}\{1 + Ca g_{ij}^{(Ca)}\}$ is the radius of curvature of both the drop interfaces.

## B. High surface Péclet number limit ($Pe_s \to \infty$)

For this limiting case we focus on obtaining the drop deformation for both kinds of imposed linear flows discussed above. Without getting involved into a detailed derivation of the same we obtain the drop deformation by applying the limit $\zeta_2 \to \infty$ to the results for the limiting case of low $Pe_s$ ($Pe_s \ll 1$). Applying the aforesaid limit to the expression given in equation (33), we obtain the final expression for the constant coefficients as

$$\lim_{\zeta_2 \to \infty}\left(L_{12}^{(Ca)}\right)_{2,0} = \frac{5}{2}, \quad \lim_{\zeta_2 \to \infty}\left(L_{23}^{(Ca)}\right)_{2,0} = 0 \quad (39)$$

As can be seen from the above expression, the coefficients shown in the above expressions are independent of shear or dilatational Boussinesq number. Henceforth no further analysis on the drop is provided for the present limiting case.

## IV. Suspension Rheology

We next move forward to obtain the effective viscosity of a dilute suspension of compound drops suspended in a linear flow, that is, a simple shear flow or a uniaxial extensional flow. The bulk rheology of the dilute suspension of compound drops is solely dependent on the interfacial dynamics of the outer drop. Hence all the flow variables are evaluated at the outer drop interface and any unnecessary mention subscripts denoting the interface of the drop is avoided. For convenience we also assume that the interfacial viscosity to be the same at either of the interfaces, $S_{12}, S_{23}$, that is $Bo_{12}^{(d)} = Bo_{23}^{(d)} = Bo_d$ and $Bo_{12}^{(s)} = Bo_{23}^{(s)} = Bo_s$. The volume averaged suspension stress for an emulsion of force free particles, as was obtained by Batchelor [56], is of the form

$$\langle\boldsymbol{\tau}\rangle = -\langle p\rangle\mathbf{I} + 2\mathbf{D}_\infty + \frac{\phi}{V_d}\mathbf{S}, \quad (40)$$

where $\phi$ is the volume fraction and the present analysis is valid for the case of a dilute emulsion, that is $\phi \ll 1$. For any quantity $\kappa$, $\langle\kappa\rangle = (1/\Delta V)\int_{\Delta V} \kappa dV$ is a volumetric average, with $\Delta V$ denoting the volume of integration. $\mathbf{S}$, in the above equation, denotes a stresslet that represents the change in total stress due to the presence of a particle (or a drop) in the flow, which alters the flow dynamics as well. This stresslet according to Batchelor can be expressed as

$$\mathbf{S} = \int_{\varphi=0}^{2\pi}\int_{\theta=0}^{\pi}\left[\frac{1}{2}\{(\boldsymbol{\tau}\cdot\mathbf{n})\mathbf{x} + ((\boldsymbol{\tau}\cdot\mathbf{n})\mathbf{x})^T\} - \frac{1}{3}\mathbf{I}\{(\boldsymbol{\tau}\cdot\mathbf{n})\cdot\mathbf{x}\} - \{\mathbf{un} + (\mathbf{un})^T\}\right]r_s^2 \sin\theta d\theta d\varphi, \quad (41)$$

where $\mathbf{x}$ in the above equation denotes the position vector and $r_s$ is the equation of the outer drop surface. For the case of an uniaxial extensional flow, the effective viscosity or the Trouton viscosity of the double emulsion can be expressed as [31]

$$\frac{\mu_{ext}}{\mu_1} = \lambda_{eff} = \langle \tau_{33} \rangle - \langle \tau_{22} \rangle = \langle \tau_{33} \rangle - \langle \tau_{11} \rangle. \tag{42}$$

After some algebraic manipulations, a more explicit expression for the non-dimensional effective viscosity, $\lambda_{eff}$, is given by

$$\lambda_{eff} = \frac{\begin{bmatrix} a_1 \eta \zeta_2^2 + \{a_2 Bo_d + a_3 Bo_s + (a_4 \alpha + a_5)\phi + (a_6 \alpha + a_7)\}\zeta_2 \\ + a_8 Bo_d^2 + a_9 Bo_s^2 + a_{10} Bo_d Bo_s + (a_{11}\phi + a_{12})Bo_d + (a_{13}\phi + a_{14})Bo_s + (a_{15}\phi + a_{16}) \end{bmatrix}}{\begin{bmatrix} b_1 \eta \zeta_2^2 + \{b_2 Bo_d + b_3 Bo_s + (b_4 \alpha + b_5)\}\zeta_2 \\ + b_6 Bo_d^2 + b_7 Bo_s^2 + b_8 Bo_d Bo_s + b_9 Bo_d + b_{10} Bo_s + b_{11} \end{bmatrix}}, \tag{43}$$

where the different constant coefficients, $a_1$-$a_{16}$ and $b_1$-$b_{11}$, in the above equation has been provided in section B of the supplementary material. The same mathematical expression for effective viscosity is also obtained for the case when there is a simple shear flow in the bulk. Thus a separate analysis for the same is not provided. It can be seen from equation (43) that surfactants along the interface of the inner drop are responsible for the presence of the quadratic terms in $\zeta_2$ in both the numerator and the denominator. In the absence of any interfacial viscosity at either of the drop interfaces we recover the result as obtained by Mandal et al. [16]. On evaluating the expression of effective viscosity [equation (43)], in the limit of $R \to 0$ with $\zeta_2 = 0$, as well as $Bo_d = Bo_s = 0$ gives us the effective viscosity of a dilute emulsion of single clean drops in the flow-field which is found to match exactly with that obtained by Taylor [57]. When the effect of the surfactants on the generation of Marangoni stress is maximum ($\zeta_2 \to \infty$), the drops starts to behave as particles. Under this situation the effective viscosity is found out to be $\lim_{\zeta_2 \to \infty} \lambda_{eff} = 1 + 5\phi/2$, which is the same as that of the Einstein viscosity formula. However, the methodology used to obtain the expression for effective viscosity as shown in equation (43) has some drawbacks. It doesn't consider the interactions in between the particles (or drops) which is possible only for an extremely dilute emulsion ($\phi \to 0$). However, with the aid of the differential effective medium approach or DEMA, as used by Pal [17], the effective viscosity of relatively concentrated emulsions can be found out. The expression of effective viscosity as presented in equation (43) can be rewritten in the following format

$$\mu_{eff} = 1 + \left[ \frac{5}{2} - \frac{3}{2} \frac{(A_1 + A_2 \lambda_2)}{(A_3 + A_4 \lambda_2 + A_5 \lambda_2^2)} \right] \phi, \tag{44}$$

where the details of the constant coefficients $A_1$ - $A_5$ are provided section C of the supplementary material. A slightly higher concentration of the suspension is obtained by addition of an incremental amount of drops which thus raises the volume fraction by d$\phi$. According to Pal [17] and Krieger & Dougherty [58], the effective increase in volume fraction of the double emulsion is expressed as d$\phi/(1-\phi/\phi_m)$, where $\phi_m$ denotes the maximum volume fraction of the drops in the bulk. The corresponding increase in the effective viscosity due to rise in the concentration is d$\lambda_{eff}$. This dimensionless effective viscosity, $\lambda_{eff}$ is thus finally determined from the following differential equation [16]

$$\frac{d\lambda_{eff}}{d\phi} = \frac{f(\lambda_{eff})}{(1-\phi/\phi_m)}, \tag{45}$$

$$\text{where } f\left(\lambda_{eff}\right) = \lambda_{eff} \left[ \frac{5}{2} - \frac{3}{2} \frac{\left\{A_1 + A_2\left(\lambda_2/\lambda_{eff}\right)\right\}}{\left\{A_3 + A_4\left(\lambda_2/\lambda_{eff}\right) + A_5\left(\lambda_2/\lambda_{eff}\right)^2\right\}} \right].$$

With the help of this differential equation, we can actually solve the effective viscosity of the emulsion corresponding to different concentrations of the same, provided the initial condition is initially stated, which for the present case is taken to be $\lambda_{eff}(\phi = 0) = 1$. This above approach has been labeled as 'model 4' by Pal [17] in his work. The above differential equation is not analytically solvable as there is no explicit relationship present in between $\lambda_{eff}$ and $\phi$. Hence a numerical approach is adopted to solve the above equation. A validation of our results with the experimental results of Pal has been shown later in the present study.

## V. Results and Discussion

### A. Drop deformation

The primary aim of this study is twofold. Our first objective is to investigate the effect of the interfacial viscosity on the deformation of the compound drop and hence the stability of a dilute double emulsion. Secondly, we aim to analyze the change in bulk rheology of the double emulsion due to variation in interfacial viscosity. We thus show the variations in both the effective viscosity and drop deformation with both $Bo_d$ and $Bo_d$. We first start our discussion on the deformation of the drop. Towards this we first show the variation of the deformation of both the outer and the inner drop as a function of $\lambda_2$. In figures 2(a)-2(d), the variation of the deformation parameter for both the outer and the inner interfaces, $D_{ext,O}$ and $D_{ext,I}$, is shown due to change in $\lambda_2$ for different values of shear and dilatational Boussinesq number ($Bo_s$, $Bo_d$) respectively. Similar results are obtained, had $\lambda_3$ been chosen instead of $\lambda_2$.

It can be seen from figures 2(a)-(d) that the deformation of the compound drop takes place in two different regimes. For the outer drop, $D_{ext} > 0$, hence the outer drop deforms into a prolate shape, while for the inner drop $D_{ext} < 0$, which suggests that the inner drop deforms to an oblate shape. A proper explanation to this nature of deformation can be provided if we analyse the fluid flow in the annular region between the inner and the outer drop. To satisfy the tangential stress balance condition, the fluid flow along the two interfaces $\left(S_{12} \& S_{23}\right)$ has to be in opposite directions. The outer drop is influenced by the presence of the imposed uniaxial extensional flow while the tangential stress balance at the outer interface requires the presence of a biaxial extensional flow in the annular region. This biaxial extensional flow leads to the deformation of the inner drop to the shape of an oblate spheroid while the outer drop sticks to its prolate shape. The deformed shape of a compound drop suspended in a uniaxial extensional flow is shown in figure 3.

To analyse the effect of both the shear and dilatational interfacial viscosity separately, we first see the effect of the shear viscosity by neglecting any influence of the dilatational counterpart $\left(Bo_d = 0\right)$. This is shown in figures 2(a) and 2(c). Later in figures 2(b) and 2(d) we neglect the influence of shear viscosity $\left(Bo_s = 0\right)$ so that the sole effect of the dilatational viscosity can be investigated. We first investigate the effect of interface viscosity for a constant bulk viscosity of either of the phases.

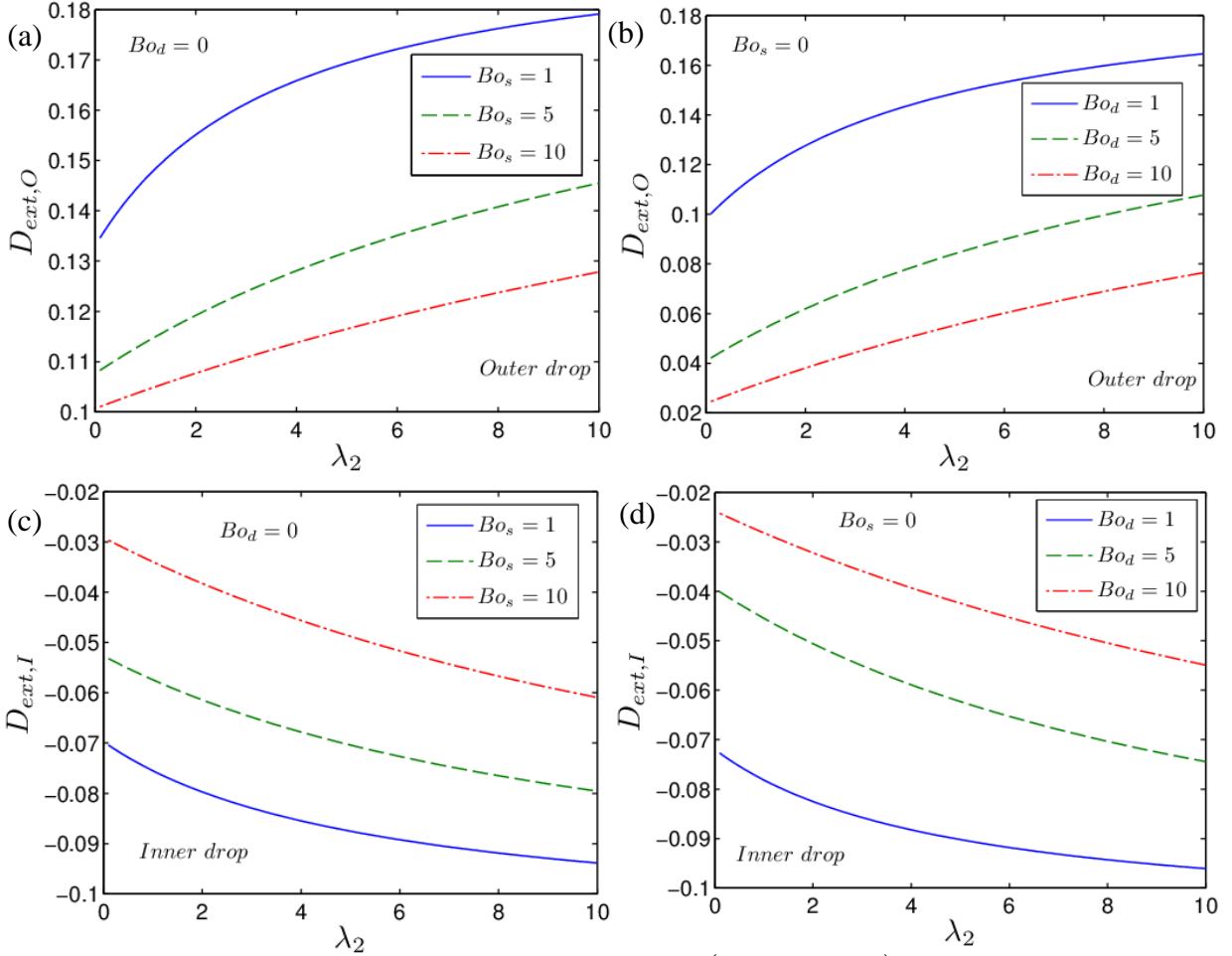

Figure 2. Variation of the deformation parameter, $D_{ext}\left(D_{ext,O}\ \&\ D_{ext,I}\right)$ with respect to $\lambda_2$ for both the outer and inner drop surfaces for different values of $Bo_s$ and $Bo_d$. Figures (a) and (c) shows the effect of $Bo_s$ on the deformation of the outer and inner drop for $Bo_d = 0$ while figures (b) and (d) shows the effect of $Bo_d$ on the deformation of the outer and inner drop for $Bo_d = 0$, respectively. The other parameters involved are $\lambda_3 = 1$, $R = 0.5$, $\eta = d = \xi = 1$, $\beta = 0.7$ and $k = 5$, $Ca = 0.1$.

Looking into the figure 2(a), we can see that increase in the interfacial shear viscosity on the interface of the outer drop leads to a decrease in its deformation ($|D_{ext,O}|$) for any fixed specified $\lambda_2$, which suggests that the outer drop tends to become more spherical from its prolate shape and hence more stable with increase in $Bo_s$. The scenario is quite similar for the case of the inner drop as well (see figure 2(c)), that is increase in $Bo_s$ results in a reduction in the deformation ($|D_{ext,I}|$) of the same which leads to a more stable spherical shaped inner drop as compared to its oblate shape. Thus the presence of shear interfacial viscosity along both the interfaces results in a more stable double emulsion. However, for a given value of $\lambda_2$ and $Bo_s$, the magnitude of deformation of the outer drop is still significantly higher as compared to the inner drop.

Moving onto figures 2(b) and 2(d), we find that the deformation of both the outer as well as the inner drop also reduces due to increase in $Bo_d$. However, there is a significant difference regarding the effect of $Bo_d$ on drop deformation as compared to that of $Bo_s$. The reduction in deformation of the drop due to increase in $Bo_d$ is relatively higher than the reduction achieved due to increase in $Bo_s$. This enhanced effect of $Bo_d$ in reducing the deformation of the drop is more significant for the outer drop as compared to that of the inner drop. Figure 3(a) and figure 3(b), shows the effect of $Bo_s$ and $Bo_d$ on the shape of the compound drop respectively. It can be clearly seen that the reduction in deformation of the outer drop due to increase in $Bo_d$ is higher as compared to that due to increase in $Bo_s$ by the same amount.

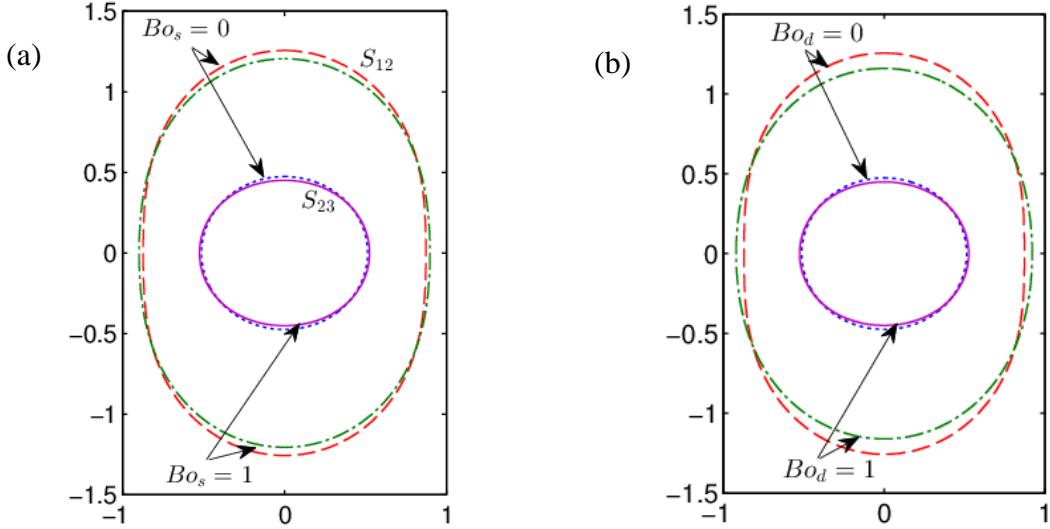

Figure 3. Shape of the compound drop shown for different values of (a) $Bo_s$ ($Bo_d = 0$). and (b) $Bo_d$ ($Bo_s = 0$). The different parameters involved in the plot are $\lambda_2 = \lambda_3 = 0, R = 0.5$, $\eta = d = \xi = 1$, $\beta = 0.7$, $k = 5$ and $Ca = 0.1$.

We now attempt to provide a physical explanation regarding the effect of interfacial viscosity on drop deformation. Towards this we rewrite the normal stress balance [equation (29)] as follows

$$\underbrace{\left[\boldsymbol{\tau}_i\big|_{S_{ij}}\cdot\mathbf{n}_{ij}-\boldsymbol{\tau}_j\big|_{S_{ij}}\cdot\mathbf{n}_{ij}\right]\cdot\mathbf{n}_{ij}}_{T_1}\underbrace{-2Bo_d\left(\nabla_s\cdot\mathbf{u}_s\right)}_{T_2}=\underbrace{\frac{\sigma_{ij}}{Ca}\left(\nabla\cdot\mathbf{n}_{ij}\right)}_{T_3}, \tag{46}$$

where $\mathbf{u}_s$ is the fluid flow velocity at the '$ij$th' interface. The part $T_1$ of the above expression denotes the jump in the normal component of hydrodynamic stress, $T_2$ signifies the normal component of the excess interfacial viscous stress solely due to the presence of dilatational viscosity along drop surface and finally $T_3$ is the capillary stress as a result of the interfacial curvature and surface tension. The extent of the drop deformation is decided by the above three terms namely $T_1$, $T_2$ and $T_3$. It may seem from equation (46) that the interfacial shear viscosity has no effect on the deformation of the drop. However, if given a closer look, it can be said that the surface shear viscosity indirectly influences the term $T_1$ through the tangential stress balance. The interfacial shear viscosity has no effect on the term $T_2$.

We first explore the influence of the interfacial viscosity on the jump in the hydrodynamic stress ($T_1$). Figure 4(a) and figure 4(b) shows the variation of the surface velocity $\mathbf{u}_s$ along the outer and the inner surface of the drop for different values of $Bo_s$ respectively. Figure 4(a) clearly indicates that the presence of interfacial shear viscosity reduces the fluid flow along the outer drop surface due to excess viscous dampening. Also according to figure 4(a), increase in the shear viscosity (or $Bo_s$) further reduces the flow strength along the surface. This reduction in the flow strength results in a decrease of the jump in the hydrodynamic stress ($T_1$) and hence the Marangoni stress responsible for the deformation of the drop reduces.

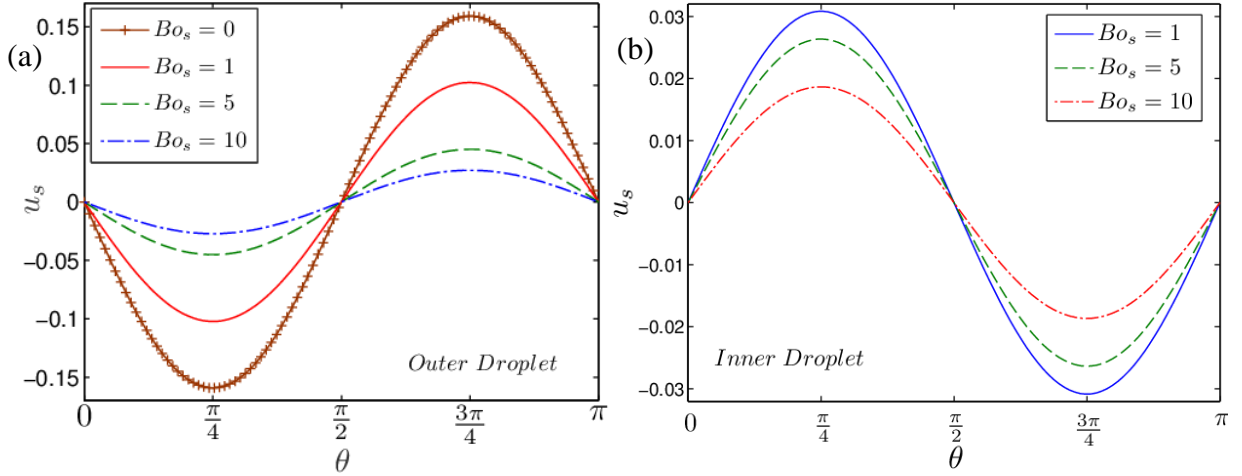

Figure 4. Variation of surface velocity of the fluid along the interface of (a) the outer drop and (b) the inner drop for different values of $Bo_s$. The other parameters involved are $Bo_d = 0$, $\lambda_2 = \lambda_3 = 1$, $R = 0.5$, $\eta = d = \xi = 1$, $\beta = 0.7$, $k = 5$ and $Ca = 0.1$.

Figure 4(b) shows the effect of the interfacial shear viscosity on the fluid flow along the inner drop surface ($S_{23}$). Firstly, it can be seen that the direction of the fluid flow is opposite to that of the outer drop which is a necessary requirement for the purpose of satisfying the tangential stress balance at the outer drop surface. Secondly, it is seen that increase in the interfacial shear viscosity also results in a decrease in the fluid flow velocity at the inner drop surface. However, the magnitude of this reduction in fluid flow is much less as compared to that for the case of the outer drop. In other words, the influence of the shear interfacial viscosity is more significant for the case of the outer interface than that of the inner surface. The reason for this difference is due to the nature of fluid flow in the annular region (biaxial extensional flow) which is different from that of the bulk (uniaxial extensional flow). Thus the interfacial shear viscosity, as a whole, enhances the stability of the double emulsion. The streamlines indicating the leading order flow field of all the three phases of the compound drop suspended in a uniaxial extensional flow in the presence of finite interfacial viscosity is shown in figure 5 for better understanding. The streamlines in the annular region appear to be squeezed by the other two bounding phases. The role played by the dilatational interfacial viscosity on the stability of the emulsion is not so straight forward. We now attempt to provide a physical explanation on the effect of the same towards deformation of a compound drop. The effect of interfacial dilatational viscosity on the fluid flow of both the interfaces ($S_{12}, S_{23}$) is quite similar to that of the interfacial shear viscosity. Presence of dilatational viscosity as well as its increase in magnitude along either of drop surfaces results in a reduction in fluid flow along the interfaces. This can be seen from both figure 6(a) and figure 6(b), where we have neglected the presence of any shear viscosity ($Bo_s = 0$) and considered only the effect of $Bo_d$. Again the effect of dilatational viscosity is seen to be less significant for the case of the inner drop as compared to the outer. This decrease in the interfacial flow which is induced due to the viscous drag reduces the net jump in the hydrodynamic stress along both the interfaces. This in turn decreases the Marangoni stress, hence reducing drop deformation.

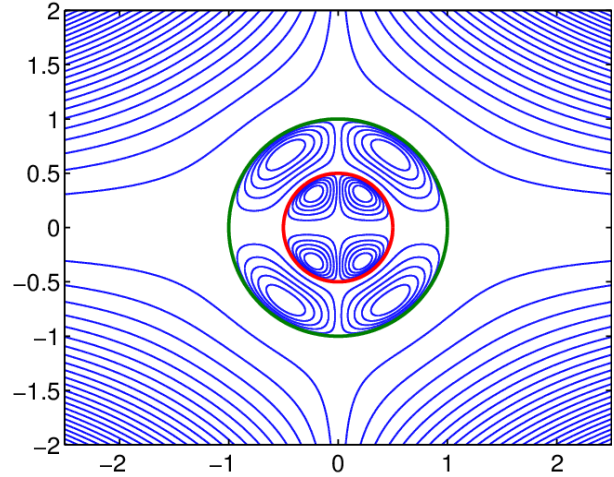

Figure 5. Contour plot of streamlines for the leading order flow field of all the three phases. The other parameter values are $\lambda_2 = \lambda_3 = 1$, $R = 0.5$, $\eta = d = \xi = 1$, $\beta = 0.7$, $k = 5$, $Bo_d = Bo_s = 5$ and $Ca = 0.1$.

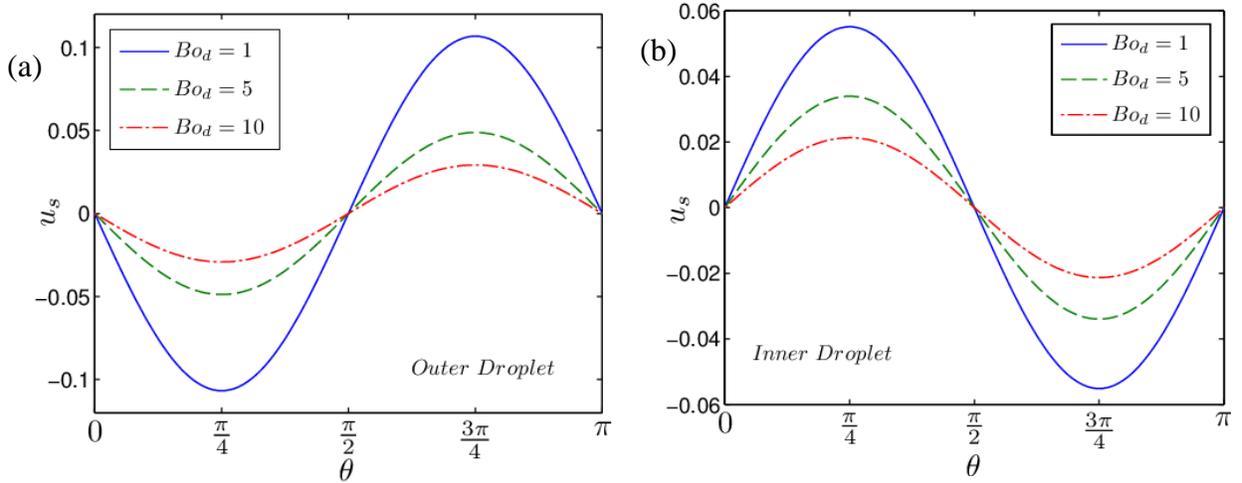

Figure 6. Variation of surface velocity of the fluid along the interface of (a) the outer drop and (b) the inner drop for different values of $Bo_d$. The other parameters involved are $Bo_s = 0$, $\lambda_2 = \lambda_3 = 1$, $R = 0.5$, $\eta = d = \xi = 1$, $\beta = 0.7$, $k = 5$ and $Ca = 0.1$.

However, the dilatational viscosity also affects the surface-excess viscous stress (the term $T_2$) that is responsible for shape deformation as well. To analyze the effect of dilatational viscosity on this excess viscous stress acting along either of the interfaces, we show the variation of the divergence of interfacial velocity ($\nabla_s \cdot \mathbf{u}_s$) along both of the drop surfaces for different values of $Bo_d$ in figures 7(a) and 7(b). As can be seen from figure 7(a), the divergence of surface velocity is positive at the equator ($\theta = \pi/2$) negative in the vicinity of the poles ($\theta = 0, \pi$). This signifies that $T_2$ is positive near the poles and is negative near the equator of the surface of the outer drop. A larger positive value of $T_2$ near the poles increases the net jump in the normal stress and hence the curvature near the poles is more.

On the contrary, a larger negative value of $T_2$ near the equator reduces the jump in the normal stress and the curvature near the equator of the interface decreases. This truly explains the prolate shape of the outer drop surface. However, our main purpose is to analyse the effect of $Bo_d$ on the deformation of the drop. It can be seen from figure 7(a) that increase in $Bo_d$ reduces magnitude of the divergence of surface velocity both near the equator as well as in the vicinity of the poles which indicates that a higher value of dilatational viscosity effectively reduces the net jump in normal stress throughout the outer drop surface. This in turn causes the reduction of curvature near the poles and increases the same near the equator resulting in a more spherical shaped drop. In other words a higher value $Bo_d$ reduces deformation of the outer drop surface. The values of the different parameters used in this plot are mentioned in the caption of figure 7.

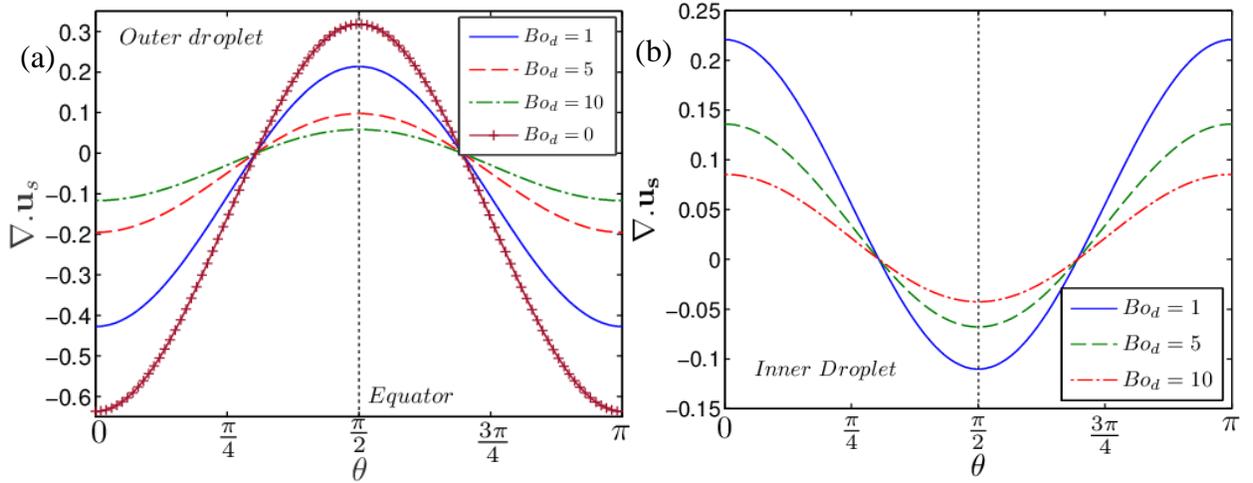

Figure 7. Variation of the divergence of surface fluid velocity with polar angle along the interface of (a) the outer drop and (b) the inner drop for different values of $Bo_d$. The other parameter values are $Bo_s = 0$, $\lambda_2 = \lambda_3 = 1$, $R = 0.5$, $\eta = d = \xi = 1$, $\beta = 0.7$, $k = 5$ and $Ca = 0.1$.

For the case of the inner drop, as shown in figure 7(b), the divergence of the surface velocity is negative near the equator and positive near the poles which suggest a positive $T_2$ near the equator and negative near the poles. This indicates a higher curvature near the equator and a lower curvature near the poles, and hence the oblate shape of the inner drop. Increase in $Bo_d$ further decreases the magnitude of $T_2$ and hence the net jump in normal stress throughout the surface of the inner drop reduces. This results in decrease in curvature near the equator and increase in the same near the poles, thus producing a more spherical shaped drop. So on one hand the dilatational viscosity decreases the drop deformation by reducing the fluid flow along the interfaces and hence the Marangoni stress. On the other hand, the dilatational viscosity reduces the contribution of the surface-excess viscous stress towards the net jump in the normal stress and decreases the deformation as well along either of the interfaces. It can be thus inferred that the dilatational viscosity plays an important role in stabilizing the emulsion and its role is more significant in comparison to that played by interfacial shear viscosity which is evitable from figures 2 and 3.

The effect of the viscosity ratio, $\lambda_2$, on drop deformation is quite obvious from figures 2(a)-(d). It is seen that for both the inner as well as outer drop, increase in the viscosity of the outer drop phase results in an enhanced deformation for any particular value of interface viscosity. Rise in viscosity is accompanied by an increase in the bulk viscous stresses which in turn results in an increased deformation of either of the interfaces. Stone and Leal [13] previously showed that the deformation of either of the interfaces of a compound drop, in the absence of any surfactants or interfacial viscosity, was enhanced due to increase in the bulk viscosity of any of the drop phases. It should however be noted that the effect of the bulk viscosity on drop deformation becomes less significant in presence of interfacial viscosity. This is especially seen for the case of low viscous drops. Looking back to figure 2, it can be seen that increase $Bo_d$ or $Bo_s$ results in a fall in the rate of increase of deformation (due to increase in $\lambda_2$) for a low viscous outer drop. Increase in interfacial viscosity results in an increase in the viscous damping which reduces the Marangoni stress responsible for deformation the drop interface. This explains the reduction in the rate of deformation. Other than this, any variation of $\lambda_2$ affects the outer drop significantly more as compared to the inner drop. This can as well be seen from figure 2. The radius ratio, $R$, also plays a similar role in drop deformation, that is increase in $R$ results in increase in drop deformation. This is because a higher $R$ signifies a larger inner drop that causes squeezing of the annular region and hence excess viscous stress develops that increases shape deformation.

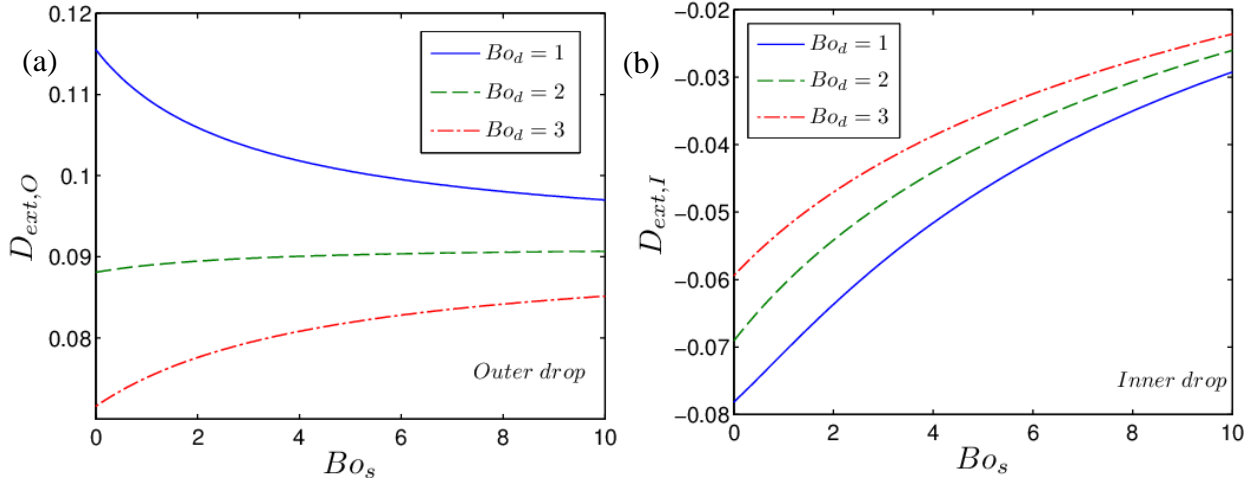

Figure 8. Plot of $D_{ext}$ for (a) the outer $(D_{ext,I})$ and (b) the inner drop $(D_{ext,O})$ as a function of $Bo_s$ for different values of $Bo_d$ respectively. The other parameter values are $\lambda_2 = \lambda_3 = 1$, $R = 0.5$, $\eta = d = \xi = 1$, $\beta = 0.7$, $k = 5$ and $Ca = 0.1$.

We next explore the combined effect of both $Bo_s$ and $Bo_d$ on the drop deformation. Towards this we show the variation of the deformation parameter for the outer and inner drop as a function of $Bo_s$ for different values of $Bo_d$ in figures 8(a) and 8(b) respectively. For the outer drop we get an interesting result. It can be seen from the plot in figure 8(a), that below a certain critical value of $Bo_d$ (= 2.2), increase in $Bo_s$, as expected, reduces the deformation of the drop thus resulting in a stable emulsion. However for $Bo_d > 2.2$, any increase in $Bo_s$ is accompanied by a corresponding rise in the deformation of the outer drop, thus rendering the emulsion unstable. From figure 8(a), it can be seen that at $Bo_d = 2$, there is almost no variation in drop deformation due to change in the shear interfacial viscosity. It should be noted that, till now, all the analysis to show the sole effect of $Bo_s$ were done under the premise of negligible interfacial dilatational viscosity ($Bo_d = 0$). Thus to preserve the stability of a double emulsion higher values of $Bo_d$ should be avoided. However, for any fixed value of $Bo_s$, increase in $Bo_d$ always reduces the deformation of the outer drop. On the other hand, for the inner drop [figure 8(b)], increase in either of $Bo_d$ or $Bo_s$ for any constant value of $Bo_s$ or $Bo_d$ always results in a reduced drop deformation, respectively. The negative sign of the deformation parameter is due to the oblate shape of the inner drop.

## B. Suspension Rheology

We now investigate on how the interfacial viscosity affects the effective viscosity of a dilute double emulsion in linear flow field. As discussed previously, the effective viscosity being same for either of the linear flows, we analyse only the case of a uniaxial extensional bulk flow. We start by presenting the variation of normalized effective viscosity $[(\lambda_{eff}-1)/\phi]$ of the emulsion of compound drops as a function of the interfacial viscosity for different size of the inner drops (for different values of $R$).

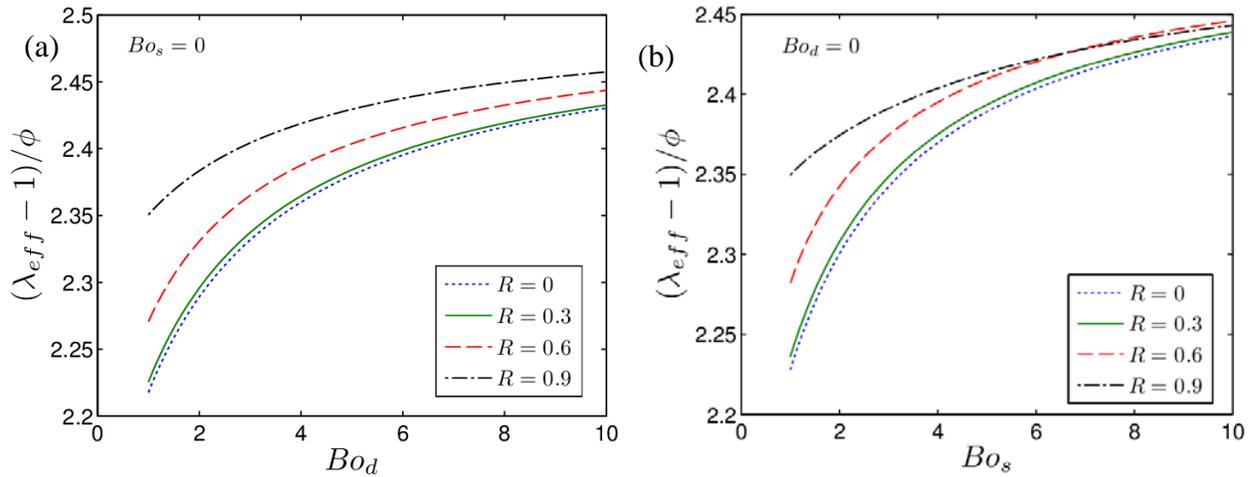

Figure 9. Variation of normalized effective viscosity $[(\lambda_{eff}-1)/\phi]$ of an emulsion of compound drops, suspended in a uniaxial extensional flow, with (a) $Bo_d$ for $Bo_s = 0$ and (b) $Bo_s$ for $Bo_d = 0$. Each of these plot are shown for different values of radius ratio, $R(0, 0.3, 0.6, 0.9)$. The other parameters involved are $\lambda_2 = \lambda_3 = 1$, $\eta = d = \xi = 1$, $\beta = 0.5$, $k = 5$ and $Ca = 0.1$.

Presence of interface viscosity, both shear and dilatational, enhances the effective viscosity of the emulsion. This can be verified from both figures 9(a) and 9(b). As discussed previously, interfacial viscosity caused viscous dampening, which reduces the fluid flow velocity along either of the interfaces. This in turn results a more uniform surfactant distribution along the drop surface and hence the Marangoni stress reduces. This explains the increase in the effective viscosity of the emulsion due to increase in the interfacial viscosity. Another interesting observation can be made from figures 9(a) and 9(b), regarding the effect of $R$. It can be seen that for any particular value of $Bo_d$, a larger inner drop or a larger value of $R$, predicts a higher effective viscosity. Larger the inner drop for a constant size of the outer drop (higher $R$), more is the annular region squeezed and hence higher are the viscous stresses developed in the annular phase. This results in an increased effective viscosity. Similar behavior is also observed when $Bo_s$ is varied. The effect of $R$ on the effective viscosity is more prominent for low values of $Bo_d$ and $Bo_s$. For higher values of interfacial viscosity, the effect of the radius on the emulsion rheology ratio reduces. This immunity of increase in effective viscosity due to alteration of $R$ is more pronounced for higher values of $Bo_s$ rather than larger values of $Bo_d$ This can be seen on comparison of figures 9(a) and 9(b). This is solely due to the fact that increased interfacial viscous dampening due to higher interfacial viscosity prevents the generation of bulk viscous stress due to increase in $R$.

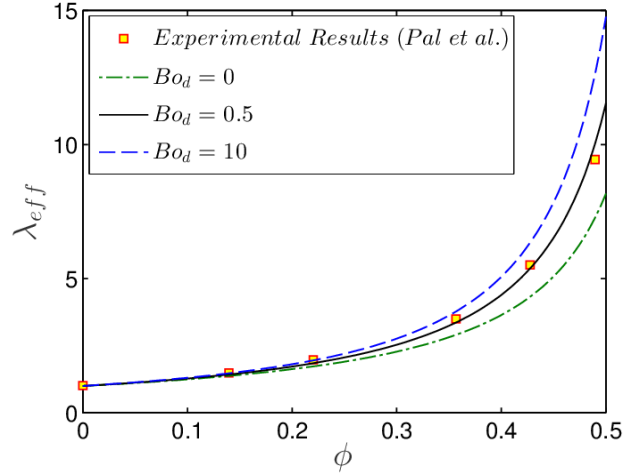

Figure 10. Variation of non-dimensional effective viscosity, $\lambda_{eff}$, with volume fraction, $\phi$, for different values of $Bo_d$. The solution for $\lambda_{eff}$ is obtained by the numerical integration of the differential equation (45). The data in the plot denoted by the 'square' marker points are those extracted from the experimental study of Pal [17]. The other parameter values are $Bo_s = 0$, $\phi_m = 0.584$, $\lambda_2 = \lambda_3 = 1$, $R = 0.5$, $\eta = d = \xi = 1$, $\beta = 0.5$, $k = 5$ and $Ca = 0.1$.

We now test the accuracy of our theoretical prediction of the effective viscosity obtained by performing numerical integration of equation (45) and comparing the results with the experimental results of Pal [17]. In the experimental study performed by Pal [17], two types of oil-water-oil double emulsions were used. The first system comprised of a mixture of water and Corna-32, while the second system was a mixture of petroleum oil and water. Experiments performed by Pal indicated that the former system displayed shear thinning behavior, even under low shear rates and low concentration. The second system, on the contrary, exhibited Newtonian behavior upto medium concentration for a wide range of applied shear rates. We thus compare the results obtained from our theoretical prediction with the experimental results for the second system. For the purpose of comparison with the experimental results of Pal we fix $k = 5$ and $\beta = 0.5$, and use $Bo_d$ as a fitting parameter with $Bo_s = 0$. Although in the above mentioned work of Pal there is no mention of any property values of the surfactant used, a later study by the same author [15] gives us the necessary details for the purpose of evaluating $k$ and $\beta$. Choosing an appropriate value for the surface diffusivity of the surfactants at the oil-water interface ($D_{12} \sim 5\times10^{-8}$ m$^2$s$^{-1}$) it can be shown that the above choice of values for $k$ and $\beta$ is justified.

Figure 10 presents a comparison between the predictions for the effective viscosity according to our theoretical model with the experimental data extracted from the results of Pal. Although there is no mention of interfacial viscosity in their study, it is quite common for a surfactant-laden interface to possess a viscosity different from the bulk. This alters our prediction on stability as well as effective viscosity of the emulsion significantly. In figure 10, we plot the non-dimensional effective viscosity, obtained by numerical integration from equation (45), against the volume fraction, $\phi$ for different values of the dilatational interfacial viscosity ($Bo_d$).

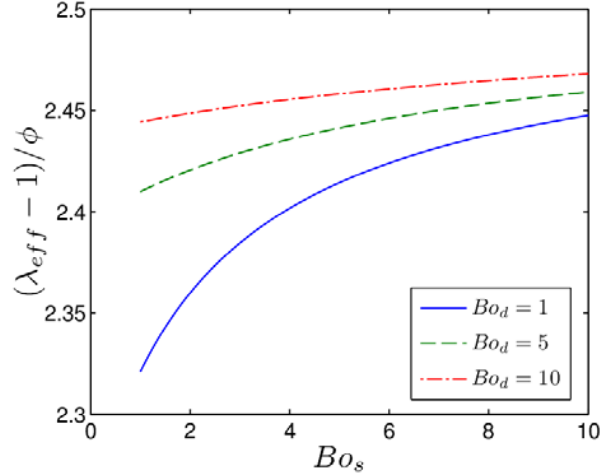

Figure 11. Plot highlighting the dependence of normalized effective viscosity [$(\lambda_{eff}-1)/\phi$] on shear interfacial viscosity ($Bo_s$) for different values of dilatational interfacial viscosity ($Bo_d$). The other parameters involved are $\lambda_2 = \lambda_3 = 1$, $R = 0.5$, $\eta = d = \xi = 1$, $\beta = 0.5$, $k = 5$ and $Ca = 0.1$.

It is seen from the figure that is a good match between both the results for $Bo_d = 0.5$, while for any other values of $Bo_d$, our theoretical prediction deviated from the experimental result. In a similar manner if we neglect the effect of dilatational viscosity ($Bo_d = 0$) and consider only the effect of the shear interfacial viscosity, it is seen that there is a good match between the results for $Bo_s = 0.6$ (not shown in figure 10). Another important observation that can be made from figure 10 is that for a high value of $\phi$ or a relatively concentrated emulsion, an increase in the shear or dilatational interfacial viscosity is accompanied by a significant increase in the effective viscosity. This can be attributed to the fact that presence of interfacial viscosity results in viscous dampening, hence reducing the flow strength along the drop interfaces and in the bulk which manifests itself by an increase in the effective viscosity. The other parameters used for the plot in figure 10 are provided in the caption.

We next look at the combined effect of both the shear and the dilatational interfacial viscosity on the effective viscosity of the double emulsion. It is seen from figure 11 that both the shear and the dilatational have similar effects on the effective viscosity of the double emulsion. Rise in either of the interfacial viscosities, results in an increased viscous drag which is accompanied by viscous dampening along the drop interfaces. This results in a decrease in the fluid flow along the interface as well as in the bulk, which in turn enhances the effective viscosity of the emulsion. It can be also seen from figure 11 that the effect of $Bo_d$ (or $Bo_s$) on the emulsion rheology is more significant for low values of $Bo_s$ (or $Bo_d$).

## VI. Conclusion

The present study deals with the role played by interfacial rheology (shear and dilatational interface viscosity) on the stability as well as on the bulk rheology of a double emulsion in an arbitrary linear flow. Both the interfaces of the compound drop is assumed to behave as a Newtonian fluid with constant interfacial viscosities and obey the Boussinesq-Scriven constitutive law. An asymptotic approach is adopted to solve the nonlinear coupled governing equations in the limiting case of diffusion dominated surfactant transport. The prediction from this study is found to have good match with previous analytical and experimental results. Some of the noteworthy outcomes of this study are stated below

  i. The deformation of both the fluid-fluid interfaces reduce as the interface viscosity (shear and dilatational) is increased. That is both shear and dilatational viscosity play their roles in stabilizing a dilute emulsion of compound drops. However, above a critical dilatational interfacial viscosity, any increase in the shear interfacial viscosity results in the rise in the deformation of the outer drop surface, hence rendering the emulsion unstable. For the inner drop surface, the deformation always reduces with increase in the dilatational viscosity irrespective of any value of the shear interfacial viscosity.
  ii. Rise in the bulk viscosity of either of the drop phases enhances the deformation of the double emulsion provided the viscosity of the carrier phase is kept constant. However presence of interfacial viscosity dilutes this effect.
  iii. Both the shear as well as the dilatational viscosity at either of the drop surface enhances the effective viscosity of the emulsion. This is due to the fact that interfacial viscosity-induced viscous dampening reduces the flow strength along the drop surface.
  iv. For the limiting case of convection dominated surfactant transport, the interfacial viscosity is found to have no significant influence on the stability as well as on the rheology of the emulsion.

**Data accessibility**: All the data used for the present study have been extracted from various published works.
**Authors' contributions**: S.D., S.M. and S.C. contributed equally to this work. All the authors gave their final approval for publication.
**Competing interests**: We declare we have no competing interests.
**Funding**: The study is not funded by any agency.

**Supplementary Material**
See the supplementary material for the details of the asymptotic analysis and the detailed expressions of the constant coefficients. The supplementary material consists of a doc file and a Matlab script file. The Matlab script file enlists the detailed expressions of different quantities which were too big to be included in the main text.